\documentstyle[preprint,aps]{revtex}

\begin{document}
\draft
                     
\title{Cosmic Rays and Gamma Ray Bursts From Microblazars} 

\author{Arnon Dar}
\address{Max Planck Institut F\"ur Physik, Werner Heisenberg Institut\\
Fohringer Ring 6, 80805 M\"unchen, Germany\\ 
and\\ 
Department of Physics and
Space Research Institute\\ 
Technion, Israel Institute of Technology, Haifa 32000,Israel}

\maketitle

\begin{abstract}

{\bf Highly relativistic jets from merger and accretion induced collapse
of compact stellar objects, which may produce the cosmological gamma ray
bursts (GRBs), are also very eficient and powerful cosmic ray
accelerators. The expected luminosity, energy spectrum and chemical
composition of cosmic rays from Galactic GRBs, most of which do not point
in our direction, can explain the observed properties of Galactic cosmic
rays.}

\end{abstract}


\newpage

The origin of high energy cosmic rays (CR), which were first discovered by
V. Hess in 1912, is still a complete mystery [1]. Their approximate broken
power-law spectrum, $dn/dE\sim E^{-\alpha}$ with $\alpha\sim 2.7$ below
the so called CR knee around $10^{15.5}~eV$, $\alpha\sim 3.0$ above the
knee and $\alpha\sim 2.6$ beyond the so called CR ankle around
$10^{19.5}~eV$, suggests different origins of CR with such energies. It is
generally believed that CR with energy below the knee are Galactic in
origin [2], those with energy above the knee can be either Galactic or
extragalactic [3], and those beyond the ankle are extragalactic [4].

If the CR accelerators are Galactic, they must replenish for the escape of
CR from the Galaxy in order to sustain the observed Galactic CR intensity. 
Their total luminosity in CR must therefore satisfy, $L_{MW}[CR]=\int
\tau^{-1}(Edn/dE)dEdV$, where $\tau(E)$ is the mean residence time of CR
with energy $E$ in the Galaxy. It can be estimated from the mean column
density, $X=\int \rho dx$, of gas in the interstellar medium (ISM) that
Galactic CR with energy $E$ have traversed.  From the the secondary to
primary abundance ratios of Galactic CR it was infered that [5]
$X=\bar\rho c\tau\ \approx 6.9 (E[GeV]/20Z)^{-0.6}~g~cm^{-2}$, where
$\bar\rho$ is the mean density of interstellar gas along their path.  The
mean energy density of CR and the total mass of gas in the Milky Way (MW),
that have been inferred from the diffuse Galactic $\gamma$-ray, X-ray and
radio emissions are [1], $\epsilon=\int E(dn/dE)dE\sim 1~eV~cm^{-3}$ and
[6] $M_{gas}=\int\rho dV\sim \bar\rho V \sim 4.8\times 10^9M_\odot$,
respectively. Hence [7],
\begin{equation}
L_{MW}[CR]\sim M_{gas}\int {cEdn/dE\over X}dE
\sim 1.5\times 10^{41}~erg~s^{-1}.
\end{equation} 

The only known Galactic sources which can supply this CR luminosity [8]
are supernova explosions [9] and gamma ray bursts [10]. Approximately,
$E_K\sim 10^{51}~erg$ is released in supernova explosions (SNe) as
nonrelativistic kinetic energy of ejecta at a rate $R_{MW}[SNe]\sim 1/30~
y^{-1}$. If a fraction $\eta \sim 15\%$ of this energy is converted into 
CR energy by collisionless shocks in the supernova remnants (SNR),  
then the total SNe luminosity in CR is,
\begin{equation}
L_{MW}[CR]\approx \eta R_{MW}[SNe]~E_K[SNe] \sim 1.5\times
10^{41}~erg~s^{-1}, 
\end{equation}
as required by eq. (1).  The non thermal X-ray emission from SNR 1006
observed by ASCA and ROSAT [11], the GeV $\gamma$-ray emission from
several nearby SNRs observed by EGRET [12], and the recent detection of
SNR 1006 in TeV $\gamma$-rays by the CANGAROO telescope [13], were all
used to argue that SNRs are the source of galactic CR. However, the TeV
$\gamma$ rays from SNRs can be explained by inverse Compton scattering of
microwave background photons by multi-TeV electrons whose synchrotron
emission explains their hard lineless X-ray radiation [11]. Furthermore,
the mean lifetime of strong shocks in SNRs limits the acceleration of CR
nuclei in SNRs to energies less than [14] $\sim Z\times$0.1PeV and cannot
explain the origin of CR with much higher energies. In fact, most nearby
SNRs in the Northern hemisphere have not been detected in TeV
$\gamma$-rays [15].  Moreover, the Galactic distribution of SNRs differs
significantly from that required to explain the observed Galactic emission
of high energy $(>100MeV)$ $\gamma$-rays by cosmic ray interactions in the
Galactic ISM [16]. All these suggest that, perhaps, SNRs are not the main
accelerators of Galacic CR.

Gamma ray bursts (GRBs) have already been proposed as CR sources
[17,18,10]. Photon acceleration of CR in isotropic GRBs is probably
limited to energies below TeV [17]. Shock acceleration in GRBs may
accelerate particles to ultrahigh energies [18].  However, because of the
assumed spherical symmetry, the total energy release in GRBs was severely
underestimated (see below). Consequently, it was concluded that
extragalactic GRBs may be the source of the ultrahigh energy CR in the
Galaxy but they cannot supply the bulk of the Galactic CR [18]. In fact,
spherical fireballs from merger and/or accretion induced collapse (AIC) of
compact stellar objects [19] cannot explain the observed properties of
GRBs and their afterglows [20]:  They cannot explain their complex light
curves and short time scale variability [21]. They cannot explain the lack
of scaling and the diversity of their afterglows [22].  They cannot
explain their delayed GeV $\gamma$ ray emission [23]. In particular,
isotropic emission implies enormous kinetic energy release in GRBs which
cannot be supplied by merger/AIC of compact stellar objects [20]. Such
release is necessary in order to sustain the long duration power-law
fading of the observed afterglow of GRB 970228 [24]. It is also needed in
order to produce the observed $\gamma$-ray fluence from GRBs 971214 and
980703 where the redshift $z$ of the host galaxy has recently been
measured [25],
\begin{equation}
E_\gamma \sim {4\pi d_L^2F_\gamma \over 1+z}{\Delta\Omega\over 4\pi}\geq
10^{53}{\Delta\Omega\over 4\pi}~erg,
\end{equation} 
where $\Delta\Omega$ is the solid angle that the emission is beamed into
and $d_L$ is their luminosity distance (we
assuming a Friedmann Universe with $h\sim 0.65$, $\Omega_m\geq 0.2 $ and
$\Omega_m+\Omega_\Lambda\leq 1$).

However, if the relativistic ejecta is beamed into a narrow jet [26], most
of the problems of the spherical fireball models can be avoided and the
main observed properties of GRBs and their afterglows can be explained
[20]. Moreover, in order to produce the same observed rate of GRBs, the
number of jetted GRBs must be larger by a factor $4\pi/\Delta\Omega$ than
the number of GRBs with isotropic emission.  Thus, for a fixed energy
release per event, the total kinetic energy release in GRBs is larger by a
factor $4\pi/\Delta\Omega$ than that which was estimated [18] for
spherical GRBs. Highly relativistic jets are also very eficient CR
accelerators. Acceleration to CR energies can take place in the jets by
diffusive shock acceleration or in front of the jets by the Fermi
mechanism [27]. In this letter I show that jets from mergers/AIC of
compact stellar objects may be the source of Galactic CR [10].  The
predicted source luminosity, energy spectrum and chemical composition
agree with those required by CR observations.

Highly relativistic jets seem to be emitted by all astrophysical systems
where mass is accreted at a high rate from a disk onto a central black
hole (BH). They are observed in galactic superluminal sources, such as the
microquasars GRS 1915+105 [28] and GRO J1655-40 [29] where mass is
accreted onto a stellar BH, and in many extragalactic blazars where mass
is accreted onto a a supermassive BH. The emission of Doppler shifted
Hydrogen Ly$\alpha$ and Iron K$\alpha$ lines from the relativistic jets of
SS443 [30] suggest that the jets are made predominantly of normal hadronic
plasma. Moreover simultaneous VLA radio observations and X-ray
observations of the microquasar GRS 1915+105 indicate that the jet
ejection episodes are correlated with sudden removal of accretion disk
material into relativistic jets [31].  Highly relativistic jets may be the
merger/AIC death throws of close binary systems containing compact stellar
objects [10,20,26]. But, because the accretion rates and magnetic fields
involved are enormously larger compared with normal quasars and
microquasars, the bulk motion Lorentz factors of these jets perhaps are
much higher, $\Gamma \sim 1000$ as inferred from GRB observations.  Such
highly relativistic jets which point in (or precess into) our direction
(`microblazars')  can produce the cosmological GRBs and their afterglows
[20]. Jetting the ejecta in merger/AIC of compact stellar objects can
solve the energy crisis of GRBs [20,25] by reducing the total inferred
energy release in GRBs by a factor $\Delta\Omega/4\pi$.  If NS-NS and 
NS-BH mergers are the triggers of GRBs [32], then beaming angles
$\Delta\Omega/4\pi \sim 10^{-2}$ are required in order to match the
observed GRB rate [21] and the currently best estimates of the NS-NS
and NS-BH merger rates in the Universe [33]. Such angles are typical of
superluminal jets from blazars and microquasars.  The estimated rate of
AIC of white dwarfs (WD) and NS in the observable Universe is $\sim 1$ per
second, i.e., larger than the estimated rate for NS-NS and NS-BH mergers
[33] by about two orders of magnitude. Therefore, if GRBs are produced by
jets from AIC of WD and NS then $\Delta\Omega/4\pi\sim 10^{-4}$.  Note
that either the 'firing' of many highly relativistic fragments into a
small solid angle [20] or precessing jets [33] can produce the complex
time structure of GRBs.
 
The energy release in merger/AIC of compact stellar objects is bounded by
$E_b\sim M_{NS}c^2\approx 2.5\times 10^{54}~erg$, where $M_{NS}\sim
1.4M_\odot$ is the gravitational mass of a typical neutron star (NS). Then
the typical kinetic energy release which is beamed into a solid angle
$\Delta\Omega$ may be of the order $E_K[GRB]\sim 2.5\times
10^{54}(\Delta\Omega/4\pi)~erg $. Such kinetic energy release was inferred
from the optical afterglows of GRBs [20]. It is also follows from the
energy release in $\gamma$-rays from GRBs with measured redshifts [25] if
the conversion efficiency of jet kinetic energy into $\gamma$-ray energy
is a few percent. The rate of SNe and cosmological GRBs {\it that point in
our direction} were estimated to be [34] $R_{L^*}[SNe]\sim 0.02~yr^{-1}$
and [35] $R_{L^*}[GRB]\sim 2\times 10^{-6}~yr^{-1}$, respectively, per
$L^*$ galaxy. If SNe and GRBs have similar histories (evolution functions) 
then the rate of GRBs in the Milky Way that point in our direction is
$R_{MW}[GRB]\sim 10^{-4}(4\pi/\Delta\Omega)R_{MW}[SNe]$. Thus, if most of
the kinetic energy released in GRBs is converted into CR energy (see
below) then the Galactic luminosity in CR due to Galactic GRBs is
\begin{equation} 
L_{MW}[CR]\sim 10^{-4}R_{MW}[SNe]E_b\sim 1.5\times 10^{41}~erg~s^{-1}, 
\end{equation} 
where the unknown GRB beaming angle has been canceled out. Note the
agreement between eq. (1) and eq. (4). 
However, the estimated ratio between the rates of SNe and GRBs is
sensitive to the choice of cosmological model, even if their histories 
were identical, This is because SNe and GRB observations
employ different techniques, have different sensitivities and consequently
sample different volumes of the Universe.  Therefore, I have estimated
$L_{MW}[CR]$ in other independent ways. For instance, I have assumed that
the ratio between the Galactic rate of GRBs and the global rate of GRBs is
equal to the ratio between the Galactic broad band luminosity [36],
$L_{MW}\sim 2.3\times 10^{10}L_\odot$, and the 
luminosity of the whole Universe, $L_{UNIV}\sim \int
(1+z)^{-1}\rho_L(z)(dV_c/dz)dz $ where $\rho_L(z)$ is the comoving
luminosity density and the factor $1/(1+z)$ is due to the cosmic time
dilation. If one assumes that most of the contribution to the volume
integral comes from redshifts where $\rho_L(z)$ is well approximated by
its measured value in the local universe [37], $\rho_L\sim 1.8h\times
10^8 L_\odot~Mpc^{-3}$ then
\begin{equation}
R_{MW}[GRB]\approx {R_{UNIV}[GRB]L_{MW}\over\rho_L\int(1+z)^{-1}(dV_c/dz)dz}. 
\end{equation} 
For a Friedmann Universe with $\Omega=1$ and $\Lambda=0$, the volume
integral yields
\begin{equation} 
\int{dV_c\over 1+z}=16\pi \left({c\over H_0}\right)^3
\int_0^\infty{(1+z-\sqrt{1+z})^2\over (1+z)^{9/2}dz}= {32\pi\over
30}\left({c\over H_0}\right)^3.  
\end{equation} 
Taking into consideration threshold and triggering effects, the estimated
rate of cosmological GRBs which point in our direction from the BATSE
observations [21] is $\sim 5~day^{-1}$. Consequently, for $h\sim 0.65$
eq. (5) yields $L_{MW}[CR]\sim 10^{41}~erg~s^{-1}$, consistent with eq. (4).
Moreover, if the `standard candle' $E_K[GRB]$ is taken to be proportional
to $<E_\gamma[GRB]>$ as obtained from eq. (3), then the  estimated CR 
luminosity becomes insensitive to the choice of the specific cosmology.

The high collimation of relativistic jets over huge distances (up to tens
of pc in microquasars and up to hundreds of kpc in AGN), the confinement
of their highly relativistic particles, their emitted radiations and
observed polarizations, all indicate that the jets are highly magnetized,
probably with a strong helical magnetic field along their axis. Magnetic
fields as strong as a few tens of $mGauss$ in the jet rest frame have been
inferred from microquasar observations [28], while hundreds of $Gauss$
were inferred for GRB ejecta. The UV light and the X-rays from the jet
(and accretion disk) ionize the ISM in front of the jet.  The swept up
ISM/jet material can be accelerated by diffusive shock acceleration in the
jet.
Alternatively, the jet magnetic field can act as a magnetic mirror and
accelerate the ionized ISM particles to high relativistic energies through
the usual Fermi mechanism [27]: 

Let us denote by $M$ the total ejected mass in an ejection episode, by
$\Gamma$ its initial bulk Lorentz factor and by $n_pm_p$ the total mass of
ionized ISM that is accelerated by the jet.  In the rest frame of ejecta
with a bulk Lorentz factor $\gamma$, the charged ISM particles move
towards the jet with energy $\gamma m_pc^2$ and are reflected back by the
transverse magnetic field in the jet with the same energy. In the observer
frame their energy is boosted to $E=\gamma^2 m_pc^2$. Moreover, each time
such a charged particle is deflected by an external magnetic field (of the
ISM or a star) back into the jet, its energy is boosted again by a factor
$\gamma^2$.  Thus, for $n$ reflections the energy of the accelerated
particle can reach $ E=\gamma^{2n} m_pc^2$ (neglecting radiation losses).
Thus a GRB jet with $\Gamma\sim 10^3$ can accelerate ISM protons to
energies up to $m_pc^2\Gamma^2\sim 10^{15}~ eV$ in a single reflection,
while two reflections can impart to them energies up to
$m_pc^2\Gamma^4\sim 10^{21}~eV$ ! For the sake of simplicity let us assume
that the $n$ multiple reflections take place simultaneously (in practice
$n\leq 2$), Let us also assume a pure hydrogenic composition (the
generalization to an arbitrary composition is straight forward).  If the
jet loses most of its energy by acceleration of the ISM and not by
hadronic collisions or radiation (because of the small hadronic cross
sections for binary collisions and radiation processes), conservation of
energy and momentum reads,
 \begin{equation}
           d(Mc^2\gamma)\approx -dn_p E;~~~E=m_p\gamma^2n.   
\end{equation} 
Consequently, for an ISM with a uniform composition 
\begin{equation}
    {dn_p\over dE}\approx {M\over m_p}{1\over 2nm_pc^2} \left[{E\over
m_pc^2}\right ]^{-2+1/2n}:~~~E<\Gamma^{2n} mc^2.  
\end{equation} 
Note that the power-law spectrum is independent of whether  
the ejecta is spherical, or conical or cylindrical. It is the same for
ions and electrons as long as losses and escape are neglected. It is also
insensitive to variations in the ISM density along the radial direction.
Under our assumed "ideal" conditions, the spectrum of accelerated
particles approaches a $dn/dE\sim E^{-2}$ shape.  Efficient acceleration
continues until either the jet becomes non relativistic, or 
disperses, or the Larmor radius of the accelerated particles in the jet
rest frame $r_L\sim 3\times10^{15}(E[EeV]/\Gamma ZB[Gauss])~ cm$
becomes comparable to the radius of the jet. Moreover, the assumption of
instantaneous acceleration is unjustified when the Larmor radius of the
accelerated particles ceases to be small compared with the typical
deceleration distance of the jet. Our simple analytical estimates can be
tested in
detailed Monte Carlo simulations. However, such simulations depend on too
many unknown jet
and ISM parameters. Instead, one may assume that the energy dependence of
the escape probabilities of CR from their accelerators is similar to that
for their escape from the Galaxy, i.e., $\tau \sim (E/ZB)^{-
0.6}$ is also valid for their escape from the CR accelerators. Then,
Galactic CR are predicted to have a power-law spectrum
with a power index $\alpha=2-1/2n+2\times 0.6~,~$ i.e.,
\begin{equation}
{dn\over dE}\sim C\left({E\over E_0}\right)^{-\alpha}~~{\rm with}~~
\alpha=~^{2.70,~~E<E_0}_{2.95,~~E>E_0}~, 
\end{equation}
where $E_0\sim m_p\Gamma^2\sim A~PeV$, with $A$ being the mass number of
the CR nuclei. These predictions agree well with the cosmic ray
observations. Because the jet emits enormous fluxes of beamed radiation in
all relevant wave lengths, the ISM in front of it must be completely
ionized, Since the escape probability of accelerated nuclei decreases like
$Z^{-0.6}$ the abundances of CR nuclei are expected to be enhanced by
approximately a factor $\sim Z^{0.6}$ compared with the ISM abundances.
Ionization potential effects are expected to be washed out in the CR
composition at high energy.  If the origin of the CR knee is the
transition from `single' to `double' reflections then there should be no
change in cosmic ray composition around the CR knee, as claimed by recent
measurements [36].
 
In conclusion, Galactic GRBs may be the main source of Galactic CR.

\end{document}